\documentclass[twocolumn,prb,
showpacs,amsfonts,amssymb,amsmath,floats,superscriptaddress,aps,longbibliography]{revtex4-2}

\usepackage{graphicx}
\usepackage{upgreek}
\usepackage{color}
\usepackage[section]{placeins}
\usepackage[colorlinks=true]{hyperref}
\usepackage[english]{babel}

\def\expect#1{\left<#1\right>}
\def\mr#1{\mathrm{#1}}

\def\up{\uparrow}
\def\down{\downarrow}

\def\sz{S^z}
\def\iz{I^z}
\def\izk{I^z_k}
\def\esz{\expect{S^z}}
\def\eiz{\expect{I^z}}
\def\eiziz{\langle\left(I^z\right)^2\rangle}
\def\eaiz{\expect{\left| I^z \right|}}
\def\esziz{\expect{S^z I^z}}
\def\eszizsziz{\langle\left(S^z I^z\right)^2\rangle}

\def\ak{A_k}
\def\a0{A_0}

\def\be{\beta_e}
\def\bn{\beta_n}

\def\we0{W_e^{(0)}}
\def\wn0{W_n^{(0)}}

\usepackage{bm}

\begin{document}

\title{Kinetic approach to the nuclear-spin polaron formation}

\author{Andreas Fischer}
\thanks{These authors contributed equally to this work.}
\address{Lehrstuhl f\"ur Theoretische Physik II, Technische Universit\"at Dortmund,
Otto-Hahn-Stra{\ss}e 4, 44227 Dortmund, Germany}

\author{Iris Kleinjohann}
\thanks{These authors contributed equally to this work.}
\address{Lehrstuhl f\"ur Theoretische Physik II, Technische Universit\"at Dortmund,
Otto-Hahn-Stra{\ss}e 4, 44227 Dortmund, Germany}

\author{Frithjof B.\ Anders}
\address{Lehrstuhl f\"ur Theoretische Physik II, Technische Universit\"at Dortmund,
Otto-Hahn-Stra{\ss}e 4, 44227 Dortmund, Germany}

\author{Mikhail M.\ Glazov}
\address{Ioffe Institute, 194021 St.Petersburg, Russia}

\date{\today}

\begin{abstract}

Under optical cooling of nuclei, a strongly correlated nuclear-spin polaron state can form in semiconductor nanostructures with localized charge carriers due to the strong hyperfine interaction of the localized electron spin with the surrounding nuclear spins.
Here we develop a kinetic-equation formalism describing the nuclear-spin polaron formation.
We present a derivation of the kinetic equations for an electron-nuclear spin system coupled to reservoirs of different electron and nuclear spin temperatures which generate the exact thermodynamic steady state for equal temperatures independent of the system size.
We illustrate our approach using the analytical solution of the central spin model in the limit of an Ising form of the hyperfine coupling.
For homogeneous hyperfine coupling constants, i.e., the box model, the model is reduced to an analytically solvable form.
Based on the analysis of the nuclear-spin distribution function and the electron-nuclear spin correlators, we derive a relation between the electron and nuclear spin temperatures, where the correlated nuclear-spin polaron state is formed.
In the limit of large nuclear baths, this temperature line coincides with the critical temperature of the mean-field theory for polaron formation.
The criteria of the polaron formation in a finite-size system are discussed.
We demonstrate that the system's behavior at the transition temperature does not depend on details of the hyperfine-coupling distribution function but only on the effective number of coupled bath spins.
In addition, the kinetic equations enable the analysis of the temporal formation of the nuclear-polaron state, where we find the build-up process predominated by the nuclear spin-flip dynamics.

\end{abstract}

\maketitle

\section{Introduction}

Intertwined dynamics of electron and nuclear spins in semiconductor quantum dots attracts increasing interest nowadays~\cite{dyakonov_book,glazov2018electron}. The hyperfine coupling of electron and nuclear spins limits the spin coherence time of the localized charge carriers~\cite{PhysRevLett.88.186802,merkulov02}, provides the dynamical polarization of nuclear spins~\cite{opt_or_book,PhysRevLett.86.5176,PhysRevLett.96.167403,PhysRevB.74.081306,PhysRevLett.99.056804}, and is responsible for the manifestations of the nuclear spins in optical response of nanosystems~\cite{artemova85,PhysRevLett.111.087603,2015arXiv150605370B}. Polarized nuclei provide a substantial effective magnetic field acting on the electron spin which amounts to several Tesla in GaAs~\cite{glazov2018electron,kkm_nucl_book,PhysRevB.101.115302,Chekhovich2017}. Also, the nuclear spin system is rather weakly coupled to the environment and nuclear spin polarization can be preserved for hours~\cite{kkm_nucl_book,2016arXiv161201699V}. It opens up prospects of using the nuclear spins in semiconductor nanosystems for various spintronics applications.

If the hyperfine interaction is sufficiently strong, it could result in the correlated state of the electron and nuclear spins.
Such a state, where the electron and nuclear spins are arranged to minimize the total hyperfine-coupling energy, is termed nuclear polaron.
The nuclear-polaron state is predicted to form under the conditions of the optical orientation in semiconductors~\cite{Merkulov:1998aa}. Within a quasiequilibrium mean field approach well developed also for bound polarons in diluted magnetic semiconductors~\cite{nagaev:polaron_eng,RS1983,Wolff:1988aa,merkulov:polaron,Merkulov:1998aa} the electron and nuclear spin systems can be characterized by effective temperatures $T_e$ and $T_n$, respectively~\cite{abragam,2017arXiv170602528V}, the latter can be positive or negative 
depending on the conditions of the dynamical nuclear polarization~\cite{opt_or_book,kkm_nucl_book}.
The notion of two different effective spin temperatures \cite{Merkulov:1998aa,PhysRevB.95.245209} assigned to weakly coupled subsystems emerges from a steady-state non equilibrium situation characterized by the dynamic distribution functions that have maintained their thermodynamic form.

If the nuclear spin temperature, $T_n$, is sufficiently low, the nuclear spins align in accordance with the fluctuating electron spin, and, in turn, support the electron spin polarization.
The mean-field approach immediately gives an estimate for the critical nuclear spin temperature for the polaron formation~\cite{Merkulov:1998aa,PhysRevB.95.245209,glazov2018electron}
\begin{equation}
\label{Tc:intro}
T_{n,c} \sim \frac{a^2}{k_B T_e} \sum_k \left| \psi (r_k) \right|^4,
\end{equation}
where $a$ is the hyperfine-coupling constant and $\psi (r_k)$ is the electron wave function at the position of a nucleus $k$. 
At typical parameters of GaAs-based systems, $T_{n,c}$ can be estimated as $10^{-7} \ldots 10^{-6}$~K depending on the electron localization volume and the electron temperature~$T_e \sim 1$~K.

Clearly, such low effective nuclear spin temperatures cannot be achieved by conventional cooling but rely on optical cooling protocols and the very weak interaction of the nuclear subsystem with its environment~\cite{kkm_nucl_book,2016arXiv161201699V}.
The experimental efforts of cooling down the nuclear spin system with the aim to observe the nuclear-spin polaron are ongoing~\cite{PhysRevLett.98.107401,2017arXiv170602528V}.
At the same time, the theory of the nuclear-spin polaron in semiconductor nanosystems is highly demanded: 
the nuclear spin fluctuations beyond the mean-field theory have been accounted for recently~\cite{PhysRevB.95.245209}.

Here we propose a kinetic model of the nuclear-spin polaron formation in nanosystems with localized charge carriers: donor-bound electrons and electrons in quantum dots.
Our calculations are based on the exact solution of the model hyperfine-interaction Hamiltonian
for which we  derive and solve the kinetic equation for the system's distribution function.
The results 
demonstrate a good agreement with the mean-field theory of the transition temperature.
Moreover, our approach correctly reproduces the fully equilibrium situation where the electron and nuclear spin temperatures are the same:
The polaron formation is described by a smooth crossover rather than by a critical phase transition.
We discuss the criteria of the polaron formation, consider the influence of the hyperfine-coupling constant distribution, and address the kinetics of the nuclear-spin polaron formation.

The paper is organized as follows.
In Sec.~\ref{sec:model}, we derive the kinetic equations by taking into account thermal reservoirs for the electron and the nuclear spins and establishing the corresponding spin-flip rates.
We start with a general electron-nuclear spin system coupled by hyperfine interaction in Sec.~\ref{sec:generalmodel} and specify the rate equations for our model Hamiltonian in Sec.~\ref{sec:simplified} which allows for the analytical considerations in Sec.\ref{sec:analytic}.
Section~\ref{sec:results} contains the results obtained by our approach.
We present steady-state spin expectation values in Sec.~\ref{sec:numeric} and compare them to mean-field calculations in Sec.~\ref{sec:mf}.
In Sec.~\ref{sec:transition} a criterion for the polaron formation is established.
The spin system with a distribution of hyperfine coupling constants is examined in Sec.~\ref{sec:ak}.
Finally, we provide results for the temporal evolution of the polaron in Sec.~\ref{sec:timeevolution}.
A conclusion of our findings is given in Sec.~\ref{sec:conclusion}.

\section{Model}
\label{sec:model}

\subsection{General analysis}
\label{sec:generalmodel}

We consider an electron spin $\bm S$ in a bath of nuclear spins with the hyperfine-coupling Hamiltonian in the form~\cite{dyakonov_book,glazov2018electron}
\begin{equation}
\label{HF:gen}
H_{hf} = \sum_{k=1}^N \sum_{\alpha\beta} A_{k,\alpha\beta} I_{k}^{\alpha} S^\beta. 
\end{equation}
Here, the individual nuclear spins are labeled $\bm I_k$ with index $k \in \{1, \ldots, N\}$, which enumerates the $N$ nuclei within the charge carrier localization volume, $\alpha$ and $\beta \in \{ x,y,z\}$ are the Cartesian indices, and $A_{k,\alpha\beta}$ is the hyperfine-interaction parameter of the $k$-th nucleus with the electron.
In this definition of the interaction constants, $A_{k,\alpha\beta}$ incorporates the electron wave function at the position of the nucleus $k$.
Equation~\eqref{HF:gen} is written in the general form and accounts for the possible anisotropy of the hyperfine interaction.
It includes the isotropic limit where $A_{k,\alpha\beta} \propto \delta_{\alpha\beta}$ with $\delta_{\alpha\beta}$ being the Kronecker $\delta$-symbol typical for the conduction band electrons in III-V and II-VI semiconductors, as well as the Ising limit where $A_{k,\alpha\beta} \propto \delta_{\alpha,z} \delta_{\beta,z}$ relevant for the valence band heavy holes in the same material systems~\cite{glazov2018electron}.
It can also account for more involved situations like in two-dimensional transition metal dichalcogenides~\cite{2018arXiv181006449A}. 

In addition to the hyperfine interaction described by Hamiltonian~\eqref{HF:gen}, electron and nuclear spins are subject to randomly fluctuating effective magnetic fields related to interactions of the spin system with its environment and also to the dipole-dipole interactions between the nuclei~\cite{glazov2018electron}.
Corresponding interactions can be represented as perturbations acting on the nuclear spins
\begin{subequations}
\label{perturb}
\begin{equation}
\label{V:nucl}
V_n = b_0\sum_{k,\alpha} b_k^\alpha I_k^\alpha,
\end{equation}
and the electron spin
\begin{equation}
\label{V:e}
V_e = B_0 \sum_{\alpha} B^\alpha S^\alpha,
\end{equation}
\end{subequations}
respectively.
Here, $b_0$ and $B_0$ are dimensional constants describing the strength of the perturbations, while the dimensionless operators $b_k^\alpha$ and $B^\alpha$ describe the effective magnetic fields acting on the corresponding spin. We assume that these operators obey standard spin-commutation relations
\begin{subequations}
\label{commute}
\begin{align}
&[b_k^\alpha,b_{k'}^\beta]= i\delta_{kk'}\epsilon_{\alpha\beta\gamma} b_k^\gamma, \\
&[B^\alpha,B^\beta]= i  \epsilon_{\alpha\beta\gamma} B^\gamma, \\
&[b_k^\alpha,B^\beta]=0. \label{commute:en}
\end{align}
\end{subequations}
with $\epsilon_{\alpha\beta\gamma}$ being the Levy-Civita symbol.
The operators corresponding to the fields acting on different nuclei as well as the fields acting on the electron and nuclei commute.
In our approach, we ignore the specific physical nature of the fields.
We, however,  assume that the effect of those ``agents'' producing the fields, i.e., neighboring nuclei in the case of the dipole-dipole interaction, or lattice phonons (via the spin-orbit coupling) can be described by the temperature $T_n$ in the case of the agents acting on the nuclei and the temperature $T_e$ for agents acting on the electron.

Let $|m\rangle$ be the eigenfunctions of the Hamiltonian~\eqref{HF:gen} and $E_m$ be its eigenenergies where $m\in\{1,\ldots, D\}$ and the Hilbert space dimension $D=(2S+1)(2I+1)^N$ for spin lengths $S$ and $I$. We introduce $f_m$, the distribution function of the coupled electron-nuclear spin system, as the diagonal part of the full density matrix $\varrho_{mm'}$, i.e., $f_m = \varrho_{mm}$. Under standard assumptions of weak perturbations $V_n$, $V_e$ the kinetic equation for $f_m$ as a function of time reads
\cite{Feynman72}
\begin{equation}
\label{kinetic}
\frac{\partial f_m}{\partial t} = \sum_{m'} (W_{mm'} f_{m'} - W_{m'm}f_m).
\end{equation}
Here $W_{mm'}$ is the transition rate from the state $|m'\rangle$ to the state $|m\rangle$ due to the action of the fluctuating fields $\bm b_k$, $\bm B$.
It can be expressed via Fermi's golden rule as the sum
\begin{equation}
\label{Wmm'}
W_{mm'} = W_{n,mm'} + W_{e,mm'}
\end{equation}
with the contributions
\begin{align}
\label{eq:fermi}
W_{i,mm'} = \sum_{r_i,r_i'\atop \Delta \epsilon = - \Delta E} &\frac{2 \pi}{Z_i} \exp \left(-\frac{{\epsilon_{r_i'}}}{k_B T_i}\right) | \langle {r_i}| \langle {m}| V_i | {m'} \rangle |{r_i'} \rangle |^2 
\end{align}
due to the perturbations $V_i$, where $i \in \{n,e\}$ distinguishes the nuclear/electronic reservoir, $\epsilon_{r_i}$ is the energy of the reservoir eigenstate $|r_i\rangle$, $\Delta \epsilon = \epsilon_{r_i} - \epsilon_{r_i'}$ and $\Delta E = E_m-E_{m'}$ denote the energy differences, and $Z_i$ is the partition function of the corresponding reservoir.
Interference contributions of the perturbations $V_e$ and $V_n$ are absent due to the commutation relations~\eqref{commute:en}.
In this approach the  nuclear spin diffusion and non-Markovian dynamics of the system are neglected.

The fact, that we describe the agents by their respective temperatures, makes it possible to relate the partial rates of the direct $m'\to m$ and the reverse $m\to m'$ processes:
\begin{subequations}
\label{balance}
\begin{align}
\frac{W_{n,mm'}}{W_{n,m'm}} = \exp{\left(\frac{{E_{m'}} - E_m}{k_B T_n} \right)},\\ 
\frac{W_{e,mm'}}{W_{e,m'm}} = \exp{\left(\frac{{E_{m'}} - E_m}{k_B T_e} \right)}.
\end{align}
\end{subequations}
If the electron and nuclear temperatures were the same, $T_n = T_e \equiv T$, the total transition rates $W_{mm'}$ and $W_{m'm}$ obey the same relation as Eq.~\eqref{balance} and the system in the steady state is described by the thermal distribution function:
\begin{equation}
\label{thermal}
f_m = Z^{-1} \exp{\left(-\frac{E_m}{k_B T}\right)},
\end{equation}
regardless of the particular values of the transition rates, where $Z$ is the partition function of the system.

\subsection{Simplified model}
\label{sec:simplified}

While for a small number of nuclei the hyperfine Hamiltonian $H_{hf}$ can be diagonalized numerically for arbitrary coupling constants \cite{PhysRevB.76.014304,Bortz2010,PhysRevB.89.045317}, it is instructive to consider a simplified model where the eigenstates can be found analytically. To that end, we focus on the particular Hamiltonian
\begin{align}
H = \sum_{k=1}^N \ak \izk \sz
\label{eq:h}
\end{align}
which is a special case of the Hamiltonian \eqref{HF:gen} with $A_{k,\alpha\beta} = \ak \delta_{\alpha,z}\delta_{\beta,z}$ and takes into account the Ising-like hyperfine interaction with the main axis being $z$.
The Hamiltonian \eqref{eq:h} fulfills the commutator relation $[H,S^z]=[H,I^z_k]=0$ , which yields the energy eigenstates in the form of direct products $|S^z\rangle |\{I^z_k\} \rangle$. Note that a 
possible physical realization of Hamiltonian~\eqref{eq:h} can be heavy-hole spins in III-V or II-VI quantum dots~\cite{glazov2018electron}. The extension of the results to the general form of the hyperfine interaction will be given elsewhere.
Also, for the sake of simplicity we set the spin length $1/2$ for all spins in the system. 
This assumption is exact for an electron spin, as well as for a bound 
hole state in a quantum dot when the $S=3/2$ spin multiplet is reduced to a Kramers degenerate pair due to symmetry reduction.
For the nuclear spins, the approximation of spin $1/2$ is justifiable based on the large number of spins constituting the nuclear spin bath.
The hyperfine-coupling constants $\ak$ determine a characteristic energy scale by the dephasing rate of the electron spin in the nuclear spin bath $\omega_h = \sqrt{ \sum_k \ak ^2}$ which is typically in the order of $1 \; \mr{ns}^{-1}$ for QD systems \cite{A.Greilich07212006} and can be up to two orders of magnitude smaller for donor-bound electrons \cite{PhysRevB.70.195340,PhysRevB.85.121303}.

By coupling the electron-nuclear spin system with Ising-type hyperfine interaction, Eq.~\eqref{eq:h}, to two thermal reservoirs, the general kinetic equation~\eqref{kinetic} for the distribution function $f(\{\izk\},\sz)$ reduces
to
\begin{align}
\frac{\partial f(\{\izk\},\sz)}{\partial t} = & f(\{\izk\},-\sz) W_{e}(\{\izk\},-\sz)\notag \\
&- f(\{\izk\},\sz) W_{e}(\{\izk\},\sz) \notag\\
&+ \sum_{k'} \left[ \right.
\begin{aligned}[t]
&f(\{I^z_1, \ldots, -I^z_{k'}, \ldots, I^z_N\},\sz) \\
& \times W_{k'}(-I^z_{k'},\sz) \\
&\left. -f(\{\izk\},\sz) W_{k'}(I^z_{k'},\sz) \right] .
\end{aligned}
\label{eq:ke}
\end{align}
Due to the spin and energy exchange with two separate reservoirs, electron and nuclear spin flips occur independently with the rates $W_{e}$ (electron) and $W_{k'}$ (nuclei).
In the framework of Hamiltonian~\eqref{eq:h} specifying the energy $E(\{\izk\},\sz) = \sz \sum_k \ak \izk$, the electron spin-flip rate Eq.~\eqref{eq:fermi} becomes
\begin{align}
\frac{W_{e}(\{\izk\},\sz)}{\we0} = 
\begin{cases}
1, \quad  E(\{\izk\},\sz) > 0,\\
\exp{\left(-\be |\sum_k \ak \izk | \right)}, \quad \mathrm{else},
\end{cases}
\label{eq:we}
\end{align}
where the flip rate for the transition to an energetically lower state is $\we0$, $\be= 1/{k_B}T_e$ is the inverse electron spin temperature.
The flip process to a state of higher energy is suppressed exponentially with respect to $\we0$ where a large inverse electron temperature or a large energy difference by the spin flip act as factors decreasing the flip probability in accordance with Eqs.~\eqref{balance}.
Analogously, we establish the rate for flipping the nuclear spin ${k'}$ as a result of coupling to the reservoir of inverse temperature $\bn{= 1/k_BT_n}$
\begin{align}
\frac{W_{k'}(\iz_{k'},\sz)}{\wn0} = 
\begin{cases}
1, \quad  A_{k'} \iz_{k'} \sz > 0,\\
\exp{\left(-\bn | A_{k'} \sz| \right)}, \quad \mathrm{else},
\end{cases}
\label{eq:wn}
\end{align}
with flip rate $\wn0$ for a downward transition in energy.
The physical origin of the flip processes and the details of the environment, that affect the rates via 
Eq.\ \eqref{eq:fermi}, remain unspecified and are aggregated in the parameters $\we0$ and $\wn0$.
Although $\we0$ and $\wn0$ do not enter the ratio of upward/downward transition, they determine the relaxation time scale of the system.

\subsection{Analytical consideration}
\label{sec:analytic}

Aiming for an analytical expression for the steady-state solution of the kinetic equations we apply the box model in the following, i.e., we put $\ak = \a0$ for all nuclei $k$.
The box model entails degeneracy of all states $|\{\izk\}\rangle$ with the same total nuclear spin $\iz= \sum_k \izk$ and thus allows for the introduction of a distribution function $g(\iz,\sz)$ that does not depend on the individual nuclear spins but on the total nuclear spin $\iz$ only.
As a result the set of kinetic equations for $g(\iz,\sz)$ strongly simplifies
\begin{align}
\frac{\partial g(\iz,\sz)}{\partial t} &= W_{e}(\iz,-\sz) \, g(\iz,-\sz)\notag\\
& \quad - W_{e}(\iz,\sz) \, g(\iz,\sz)\notag\\
& \quad +\sum_{j=\pm1} W_{n}^{(j)} (\iz-j,\sz) \, g(\iz-j,\sz)\notag\\
& \quad - \sum_{j=\pm1} W_{n}^{(j)}(\iz,\sz) \, g(\iz,\sz),
\label{kinetic:g:box}
\end{align}
and becomes analytically solvable. The spin-flip rates for the box model read
\begin{subequations}
\label{kinetic:w:box}
\begin{align}
&\frac{W_{e}(\iz,\sz)}{\we0} =
\begin{cases}
1, \quad \a0 \sz \iz >0,\\
\exp{\left(-\be \a0 |\iz| \right)}, \quad \mathrm{else},
\end{cases}\\
\label{kinetic:wn:box}
&\frac{W_{n}^{(\pm1)}(\iz,\sz)}{\wn0} = N_\mp(\iz) \times
\begin{cases}
1, \quad \mp \a0 \sz>0,\\
\exp{\left(- \frac{\bn \a0}{2} \right)}, \quad \mathrm{else},
\end{cases}
\end{align}
\end{subequations}
where we distinguish between a nuclear spin flipping up $(+1)$ and a nuclear spin flipping down $(-1)$ and absorb the number $N_{-(+)}(\iz)$ of nuclear spins in spin down (up) state into the nuclear spin-flip rate.

Furthermore we can exploit the fact that typically the electron spin-flip rate is much larger than the nuclear spin-flip rate, $\we0 \gg \wn0$, due to the much stronger electron-lattice coupling~\cite{glazov2018electron}.
Considering the nuclear spins as frozen on the time scale of electron spin flips, the instantaneous steady-state occupation of electron spin states for a fixed nuclear spin state is given by the Boltzmann distribution
\begin{align}
\frac{g(\iz,\up)}{g(\iz,\down)} = \exp \left( -\be \a0 \iz \right) .
\label{eq:bm}
\end{align}
Subsequently the distribution of nuclear spin states can be calculated rigorously by condensing
\begin{equation}
g(\iz,\up) + g(\iz,\down) = g(\iz),
\end{equation} 
and inserting the expressions of spin-flip rates, Eqs.~\eqref{kinetic:w:box}, in the kinetic equation, Eq.~\eqref{kinetic:g:box}, with the result 
\begin{equation}
\frac{\partial g(\iz)}{\partial t} = \sum_{j=\pm1} \Gamma^{(j)}(\iz-j) g(\iz -j) - \Gamma^{(j)}(\iz) g(\iz) .\label{rate:g:simple}
\end{equation}
Here we introduced the effective nuclear spin-flip rates $\Gamma^{(\pm)}(I^z)$.
These nuclear spin-flip rates are combined into the expression
\begin{align}
\frac{\Gamma^{(\pm1)}(\iz)}{W_n^{(0)}} = N_\mp(\iz) \frac{\cosh(\be \a0 \iz /2 \pm \bn \a0/4)}{\cosh(\be \a0 \iz/2) \exp(\bn \a0/4)}
\end{align}
that contains all contributions according to Eq.~\eqref{kinetic:wn:box} for either a nuclear spin flipping up, $(+1)$, or down, $(-1)$.
Note that in the kinetic equation for $g(\iz)$ the contributions of $g(\iz,\up)$ and $g(\iz,\down)$ to the electron spin-flips following Eq.~\eqref{kinetic:g:box} exactly cancel each other.
Therefore, only the nuclear spin-flip terms remain in Eq.~\eqref{rate:g:simple}.

The steady-state solution, $\partial_t g(\iz) = 0$, is determined from the detailed balance condition 
\begin{equation}
g(\iz)\Gamma^{(+1)}(\iz) = g(\iz+1) \Gamma^{(-1)}(\iz+1),
\end{equation}
for the exchange between neighboring nuclear spin states $\iz$ and $\iz+1$ leading to the ratio
\begin{align}
\frac{g(\iz+1)}{g(\iz)} = &\frac{N/2-\iz}{N/2+\iz+1} \times \frac{\cosh\left( \be \a0 (I^z+1) / 2\right)}{\cosh\left( \be \a0 I^z / 2 \right)} \notag \\
& \times \frac{\cosh\left(\bn \a0 / 4 + \be \a0 I^z / 2\right)}{\cosh \left( \bn \a0/4 - \be \a0 (I^z+1)/2\right)} ,
\label{eq:db}
\end{align}
with the normalization condition $\sum_{\iz} g(\iz) = 1$ and $\iz = -N/2, \ldots, N/2$.
As a result, we obtain the steady-state distribution function $g(\iz,\sz)$ and calculate the observables, see Sec.~\ref{sec:results} for results.

\section{Results}
\label{sec:results}

\subsection{Numerical results}
\label{sec:numeric}

The approach via kinetic equations, as described in Sec.~\ref{sec:analytic}, enables us to calculate temperature-dependent spin expectation values and study the formation of a nuclear polaron at sufficiently low temperatures.
The expectation value of a general observable $O$ can be represented as
\begin{align}
\expect{O} = \sum_{\sz} \sum_{\{\izk\}} \left< \sz , \{\izk\} \left| O \right| \sz , \{\izk\} \right> f(\{\izk\},\sz) .
\label{eq:expect}
\end{align}
in the steady state.
Within the box model Eq.~\eqref{eq:expect} simplifies to a sum over the total spin $\iz$.

\begin{figure}[t!]
\centering
\includegraphics[scale=1]{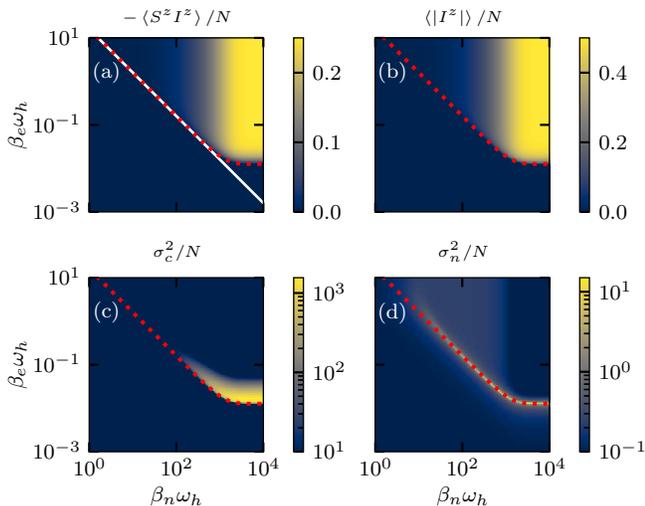}
\caption{(a) Electron-nuclear spin correlator, (b) average absolute value of the nuclear spin polarization, (c) fluctuations of the correlator, and (d) fluctuations of the absolute value of the nuclear polarization ($N=10^5$). The white solid line in panel (a) corresponds to the mean-field critical temperature, Eq.~\eqref{eq:mf-ct}. The red dotted line marks the transition to the polaron formation, Eq.~\eqref{eq:trans1}.}
\label{fig:cm}
\end{figure}

One particularly important quantity to reveal the spin orientation in the system and reflect the polaron formation is the electron-nuclear spin correlator $\esziz$, see Fig.~\ref{fig:cm}(a).
While for high spin temperatures
the $z$-components of the electron spin and the total nuclear spin
are uncorrelated ($\esziz=0$), the electron and nuclear spins align oppositely (at $\a0>0$) when the system is cooled down.
For temperatures low enough, the electron-nuclear spin correlator normalized by the number of nuclear spins finally reaches its maximum absolute value of one quarter.
Note that in Fig.~\ref{fig:cm} we use inverse temperatures, $\be$ and $\bn$,
on a logarithmic scale for illustrative purposes.
The dependence of $\esziz/N$ on electron and nuclear spin temperatures, however, is asymmetric.
For $\bn \gg \be$ the crossover from a disordered state of the system to a polaron state is very rapid but it turns smooth when $\bn$ and $\be$ become comparable in magnitude.

\begin{figure}[t!]
\centering
\includegraphics[scale=1]{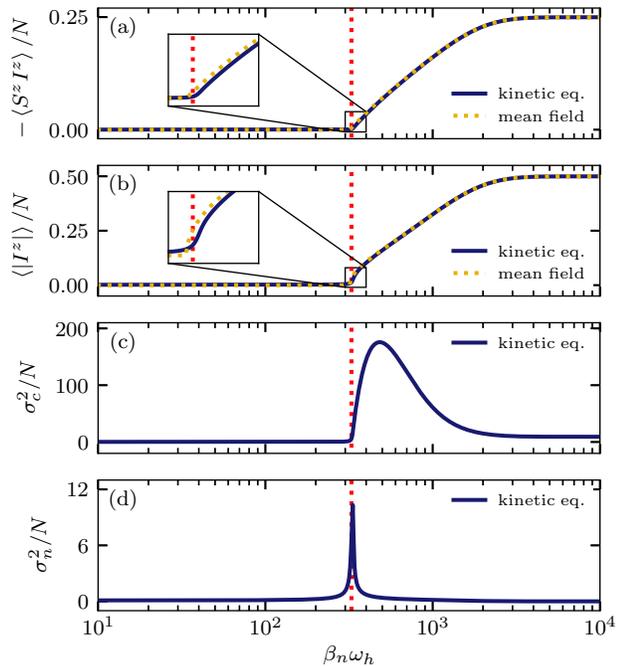}
\caption{(a) Electron-nuclear spin correlator, (b) average absolute value of the nuclear spin polarization, (c) fluctuations of the correlator, and (d) fluctuations of the absolute value of the nuclear polarization for a fixed electron temperature ($\be \omega_h =0.05$, $N=10^5$). Solid dark blue lines present results for the kinetic equations. The mean-field results, Eqs.~\eqref{mf:self}, are added as dotted orange lines. The red dotted vertical lines indicate the transition temperature for the polaron formation, Eq.~\eqref{eq:trans1}.}
\label{fig:fbe}
\end{figure}

A quantity behaving similar to the electron-nuclear spin correlator is the expectation value of the absolute value of the total nuclear spin $z$ polarization $\eaiz$, see Fig.~\ref{fig:cm}(b).
We introduce $\eaiz=\sum_{\iz} \left| \iz \right| g(\iz)$ as the absolute value of $\iz$ weighted by the nuclear distribution function $g(\iz)$ following Eq.~\eqref{eq:expect}.
The average absolute value of $I^z$, $\eaiz$, can also be interpreted as the nuclear spin polarization $\eiz_+$ of the symmetry broken distribution function 
\begin{equation}
g^{(+)}(\iz) = \theta(\iz)[g(\iz)+g(-\iz)],
\end{equation}
 and is usually used to study the polarization of a finite system, where a spontanious symmetry breaking is naturally absent $(\eiz=0)$ \cite{PhysRevB.95.245209,Binder1981,landau_binder_2014}.
While $\esziz$ and $\eaiz$ appear similar at first glance, their quantitative behavior differs.
By displaying the quantities as a function of the inverse nuclear spin temperature for a fixed electron spin temperature, this difference becomes visible [see Figs.~\ref{fig:fbe}(a) and \ref{fig:fbe}(b)].
When the polaron formation sets in, the correlator shows a nearly linear growth [see inset, Fig.~\ref{fig:fbe}(a)].
In contrast the absolute value of the total nuclear spin-$z$ component displays a square-root like behavior [see inset, Fig.~\ref{fig:fbe}(b)].
This behavior of $\esziz$ and $\eaiz$ can also be extracted from mean-field calculations \cite{Merkulov:1998aa} (see Sec.~\ref{sec:mf}).

The difference of the quantities $\esziz$ and $\eaiz$ transfers
to their fluctuations
\begin{align}
\sigma_c^2 &= \eszizsziz - \esziz^2, \\
\sigma_n^2 &= \eiziz - \eaiz^2,
\end{align}
see Figs.~\ref{fig:cm}(c), \ref{fig:cm}(d), \ref{fig:fbe}(c), and \ref{fig:fbe}(d).
The fluctuations of $\esziz$, depicted in Fig.~\ref{fig:cm}(c), exhibit a rather broad peak in the temperature range where the correlation between electron and nuclei grows.
For $\eaiz$, the fluctuations display a sharp peak, when polaron formation sets in.
In the $(\be,\bn)$ plane, this peak of $\sigma_n^2$ clearly indicates a line separating the disordered state of the system from the polaron state.
For interpretation of these fluctuations we refer to the case of a single temperature.
Here the average $\esziz$ is proportional to the energy of the system.
Thus, in thermal equilibrium the fluctuations of the electron-nuclear spin correlator define, up to a temperature-dependent prefactor, the heat capacity of the system.
Analogously, the fluctuations of $\eaiz$ can be connected to an effective finite-system susceptibility introduced in Refs.~\cite{Binder1981,landau_binder_2014}.
Note that, while the fluctuations of $\eiz$ are proportional to the ``true'' nuclear spin susceptibility, the fluctuations of $\eaiz$ are proportional to the susceptibility corresponding to the symmetry ``broken'' distribution function $g^{(+)}(\iz)$. 
According to the Landau theory of phase transitions, at the critical temperature the heat capacity would display a steplike behavior and the susceptibility a divergence~\cite{ll5_eng}.
Since we consider a finite system here, we do not have a phase transition, and thus do not observe this clear-cut behavior; rather we observe a relatively sharp cross-over as expected from the general theory~\cite{Binder1981}.

\subsection{Mean-field approach}
\label{sec:mf}

It is instructive to compare the results obtained above with the basic mean-field approach~\cite{Merkulov:1998aa}.
As a basis for the comparison, we briefly recap the mean-field calculation here.
Electron and nuclear spin expectation value are determined by the mean nuclear/electron polarization, respectively,
\begin{subequations}
\label{mf:self}
\begin{align}
\esz &= - \frac12 \tanh \left( \frac{\be \a0 \eiz}{2} \right), \label{eq:mfs}\\
\eiz &= - \frac{N}{2} \tanh \left( \frac{\bn \a0 \esz}{2} \right),
\label{eq:mfi}
\end{align}
\end{subequations}
where the effective electron and nuclear spin temperature differ.
Inserting one into the other yields a self-consistent equation that provides a nontrivial solution ($\esz \neq 0$) for a specific range of the electron and nuclear temperatures.
The mean-field approach predicts the nuclear-polaron formation once the product of the temperatures of both subsystems is below a critical constant
\begin{align}
 T_e T_n \le \frac{N \a0^2}{16 k_B^2} ,
\label{eq:mf-ct}
\end{align}
in agreement with the estimate given in the introduction, Eq.~\eqref{Tc:intro}.
For comparison with the kinetic approach, this critical line of temperatures is added in Fig.~\ref{fig:cm}(a) as a white line.
The ``critical'' behavior of the electron and nuclear spin polarizations in the vicinity of the critical temperature is predicted  by the mean-field model as
\begin{equation}
\langle |S^z|\rangle, \langle |I^z|\rangle \propto \sqrt{\be \bn -\frac{16}{N \a0^2}},
\label{eq:mf-sziz}
\end{equation}
while their values in the whole range of temperatures can be determined by numerical solution of Eqs.~\eqref{mf:self}.

We present the functional dependency of the expectation values $\esziz$ and $\eaiz$ and their fluctuations on the inverse temperature $\bn$ of the nuclear spin subsystem for a fixed  electron spin temperature of $\be \omega_h =0.05$ in Fig.~\ref{fig:fbe}.
Overall, the mean-field expectation values (orange dotted lines) are congruent with the kinetic results, though the kinetic results are smoothed at the edge due to thermal fluctuations in the finite system in the vicinity of the critical mean-field temperature. Therefore, the mean-field expectation values according to Eq.~\eqref{eq:mf-sziz} allow analytical understanding of the differing temperature dependencies of $\esziz$ and $\eaiz$ observed in Fig.~\ref{fig:fbe}.

\subsection{Criterion for the polaron formation}
\label{sec:transition}

\begin{figure}[t!]
\centering
\includegraphics[scale=1]{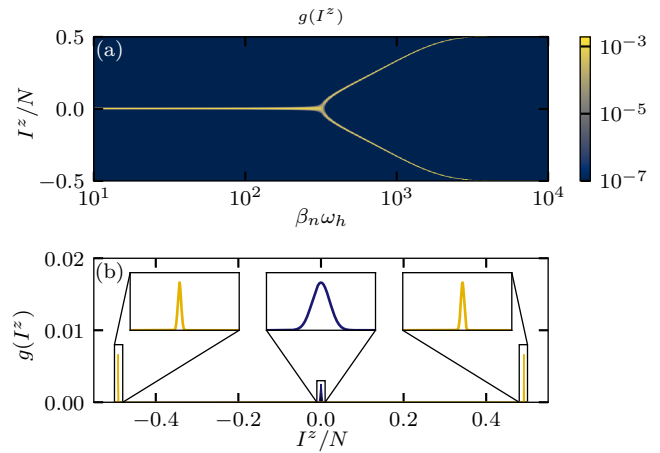}
\caption{Probability distribution $g(\iz)$ of total nuclear spin $\iz$ in steady state ($N=10^5$, $\be \omega_h = 0.05$). The upper panel displays the $\bn$ driven transition from one maximum (at $\iz=0$) to two maxima that move further apart when the system is cooled down. The lower panel presents $g(\iz)$ for $\bn \omega_h = 10$ (dark blue) and $\bn \omega_h = 3000$ (orange) as an example.}
\label{fig:giz}
\end{figure}

Although Figs.~\ref{fig:cm} and \ref{fig:fbe} provide visual indicators for the development of an antiparallel electron-nuclear spin orientation in a cooled system,
we aim for an analytic criterion where the formation of a nuclear polaron sets in.
In case of a disordered spin system the nuclear distribution function $g(\iz)$ exhibits a single maximum at $\iz = 0$, see Fig.~\ref{fig:giz}.
In contrast, when the nuclear-polaron formation starts $g(\iz)$ develops a minimum at $\iz = 0$ with two maxima $I_z \ne 0$ placed symmetrically around it.
With decreasing the nuclear spin temperature these two maxima are spaced further apart until finally the maximum alignment of nuclear spins is reached.

The transition point to the polaron formation for a fixed electronic temperature can be naturally related to the nuclear spin temperature at which the two peaks in the distribution function are starting to be formed.
We use the ratio $g(1)/g(0)$ (for even $N$) as a rigorous mathematical criterion 
for the formation of a polaron state and define the crossover line via 
\begin{eqnarray}
\frac{g(1)}{g(0)} & =& 1
\end{eqnarray}
since the splitting of the peak requires $g(1)/g(0) \geq 1$.
Using the analytical result of Eq.~\eqref{eq:db} yields the implicit condition
\begin{align}
1 =  &\frac{N}{N+2} \frac{ \cosh\left(\bn \a0 / 4 \right) \cosh\left(\be \a0 / 2\right) }{\cosh\left(\bn \a0 / 4 - \be \a0 / 2\right) } 
\label{eq:trans1}
\end{align}
for the transition line to the polaron formation.
We added this line to all panels of Fig.~\ref{fig:cm} as a red dotted line.
It coincides with the maxima of the fluctuations $\sigma_n^2$ as shown in Fig.~\ref{fig:cm}(d).

We supplemented the location of the nuclear transition temperature as vertical red dotted lines to all panels of Fig.~\ref{fig:fbe} for a fixed electron spin temperature.
For $\esziz$ and $\eaiz$ the transition temperature indicates where the expectation values start to increase substantially.
This behavior can be directly related to the emergence of peaks in $g(\iz)$ at $I^z \neq 0$. 

In the limit of small inverse electron temperatures, $\be \a0 \ll 1$, we can rewrite Eq.~\eqref{eq:trans1} to the explicit expression
\begin{align}
\bn = \frac{4}{\a0} \mr{artanh} \left( \frac{4}{(2+N) \be \a0}\right).
\label{eq:trans2}
\end{align}
This asymptotics matches the results of Eq.~\eqref{eq:trans1} within the numerical uncertainty for the chosen set of parameters.
Note that the artanh$(x)$ diverges for $|x|=1$ defining a lower bound for $\be$ for a transition to occur: $\be > 4/(2+N) \a0$.

Generally, the temperature course of the transition to a polaron state can be understood in the following way:
For a higher electron spin temperature the nuclear spins have to be cooled down further in order to achieve a polaron state.
When the electron spin temperature exceeds the upper bound $(2+N)\a0/ 4k_B$,
however, a polaron cannot form even for minimum nuclear spin temperature due to fluctuations of the electron spin.
The latter effect is absent in the mean-field approach which causes a divergence of the transition line and the mean-field critical temperature, white line in Fig.~\ref{fig:cm}(a).
It is noteworthy that for a sufficiently large number of nuclear spins (or low electron spin temperatures) where 
\[
N \gg (\be \a0)^{-1},
\]
the condition Eq.~\eqref{eq:trans2} crosses over to the mean-field critical temperature in Eq.~\eqref{eq:mf-ct}.

\subsection{Role of the distribution of hyperfine-coupling constants}
\label{sec:ak}

\begin{figure}[t!]
\centering
\includegraphics[scale=1]{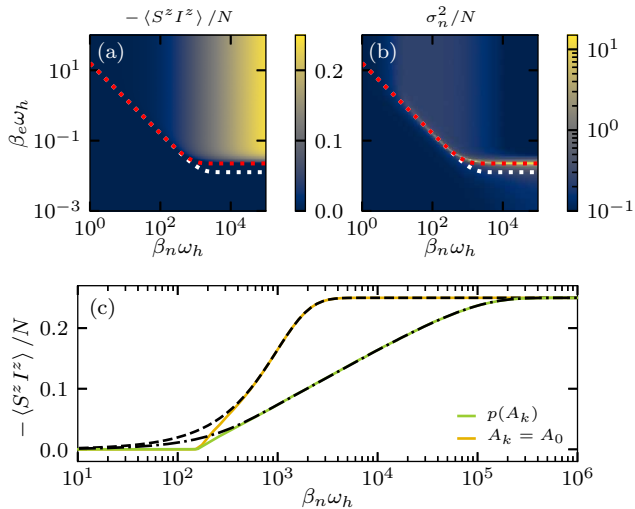}
\caption{Polaron formation in a system ($N=10^5$) with realistic hyperfine-coupling constants, see Eq.~\eqref{eq:pak}. (a) Electron-nuclear spin correlator, and (b) fluctuations of the absolute value of the nuclear spin polarization as a function of $\be$ and $\bn$. The red dotted line indicates the transition temperature, Eq.~\eqref{eq:trans1}, adjusted with $N_\mr{eff}$. The white dotted line presents the transition line with $N$. (c) Correlator for a fixed electron temperature ($\be \omega_h =0.1$). The black dashed/dash-dotted lines correspond to the approximation in Eq.~\eqref{eq:szizapprox}.}
\label{fig:cmak}
\end{figure}

Since in a real quantum dot or donor-bound electron system the hyperfine-coupling constants are determined by the electron wave function at the position of the nuclei, we lift the restriction to a fixed hyperfine-coupling constant $\a0$ in the following and take into account a realistic distribution of $\ak$.
To this end, we assume the electron envelope wave function 
$\psi (\vec{r}) \propto \exp \left[ - r^m / (2 L_0^m) \right]$ with a characteristic size of the quantum dot $L_0$. 
We use the wavefunction in this form to calculate the distribution function for hyperfine-coupling constants following Ref.~\cite{hackmann_pssb2014}
\begin{align}
p(\ak) = \frac{d}{m} \frac{L_0^d}{\ak R^d}\left[\ln\left(\frac{A_\mr{max}}{\ak}\right)\right]^{d/m - 1} \label{eq:pak}
\end{align}
where $d$ is the dimension of the quantum dot (e.g., $d=2$ corresponds to a flat dot and $d=3$ to a spherically symmetric one).
The parameters $A_\mr{max}$ and $R$ are needed to regularize the distribution function: $A_\mr{max}$ is the largest coupling constant in the center of the quantum dot and the cutoff radius $R$ determines the smallest coupling constant and regularizes the distribution $p(\ak)$.
Under the relevant assumptions of a Gaussian wave function and a flat dot, $m=2$ and $d=2$, the distribution function becomes $p(\ak) \propto 1/\ak$. 
In our calculations, we set the cutoff $R = 2.5 L_0$ and adjust $A_\mr{max}$ to provide the dephasing rate $\omega_h$.
Figure~\ref{fig:cmak} displays the steady-state results of the kinetic equations, Eq.~\eqref{eq:ke}, in a system ($N=10^5$) with $\ak$ distributed accordingly.
To gain these data we employed a Monte Carlo simulation implementing spin flips according to the flip rates in Eqs.~\eqref{eq:we} and \eqref{eq:wn}, and set $\we0 / \wn0 = 10^5$.
 
We observe that the polaron formation is shifted to lower temperatures in comparison to the box model, i.e., the anticorrelation between electron and nuclei, see Fig.~\ref{fig:cmak}(a), builds at larger $\be$ and $\bn$. 
Note that the presented temperature range is expanded compared to Fig.~\ref{fig:cm} in order to include the area with maximum correlator.
The shift to lower temperatures is reflected in the temperature line indicated by the fluctuations of $\eaiz$ as well, see Fig.~\ref{fig:cmak}(b).
We find that the initial transition line of the box model,
added as a white dotted line to the Figs.~\ref{fig:cmak}(a) and \ref{fig:cmak}(b), 
does not match the maxima line of $\sigma_n^2$.

The cutoff radius $R$ in Eq.~\eqref{eq:pak} is arbitrary and can be send to infinity in a real system while keeping $A_\mr{max}$ fixed.
That implies that the theory would include many nuclear spins that do not couple to the electron spin in the relevant energy or time window
while in the box model all nuclear spins contribute equally to the dynamics.
In order to quantify this effect and relate general results to the box model, we define an effective number of nuclear spins $N_\mr{eff}$ relevant for the dynamics \cite{0953-8984-15-50-R01,PhysRevB.98.085202}
\begin{align}
\frac{N_\mr{eff}}{N} = \frac{\left< \ak \right>^2}{\left< \ak^2 \right>}
\end{align}
by taking into account the first two moments of the distribution function $p(\ak)$.
While the box model yields $N=N_\mr{eff}$, we obtain $N_\mr{eff} / N \approx 0.32$ for the distribution function in Eq.~\eqref{eq:pak}.
Entering this effective bath size into Eq.~\eqref{eq:trans1} yields a corrected transition line (red dotted line) which coincides with the peak of $\sigma_n^2$. 
Hence, for the shape of the transition line, the effect of distributed coupling constants is reflected by the effective bath size $N_\mr{eff}$ whereas the explicit distribution function $p(\ak)$ plays a minor role.

Apart from the transition line, the distributed $\ak$ effect the system's behavior below the transition temperature.
Figure~\ref{fig:cmak}(c) presents the electron-nuclear spin correlation as a function of the inverse nuclear temperature for a fixed electron spin temperature, $\be \omega_h = 0.1$.
The green line was obtained for the hyperfine-coupling distribution $p(\ak)$, and
the box-model result is depicted as an orange line.
Compared to the box-model calculation, the maximum anti-correlation (i.e., the maximum absolute value of $\langle S^zI^z\rangle/N$ of minus one quarter) is reached for much lower nuclear temperatures.
Analytically we can shed some light into this behavior at low temperatures by the assumption of a frozen electron spin, since the electron spin flips are suppressed due to the high energy difference by a flip for strong Overhauser fields.
For a frozen electron spin, the system factorizes, and the correlator
\begin{align}
\esziz \approx - \sum_k \tanh\left( \ak \bn / 4 \right) / 4
\label{eq:szizapprox}
\end{align}
is given by the contributions of the individual nuclei.
This approximation is added to Fig.~\ref{fig:cmak}(c) for the box model (black dashed line) and the distribution in Eq.~\eqref{eq:pak} (black dash-dotted line). 
While this approximation fails close to the transition temperature,
where the assumption of a frozen electron spin does not hold and the physics is dominated by quantum fluctuations,
it precisely predicts the correlation at low nuclear temperatures.
Hence, the distribution function $p(\ak)$ governs the slope of the anti-correlation function when cooling the nuclear spin system
while the effective number of nuclear spins $N_\mr{eff}$ determines the transition point.

\subsection{Kinetics of the nuclear-polaron formation}
\label{sec:timeevolution}

In addition to the properties of the steady-state distribution function,
our model also allows to investigate their temporal evolution and address the kinetics of the polaron-state formation.
For this purpose we fix the temperatures in the polaron regime and consider the time evolution determined by the kinetic equations, Eq.~\eqref{eq:ke}.
Knowing that a distribution of hyperfine-coupling constants can be mapped to the box model by $N_{\rm eff}$,
we return to fixed coupling constants $\ak=\a0$ in this section when solving the coupled differential equations \eqref{kinetic:g:box} with finite flip rates, $\wn0$ and $\we0$.
Including $N=10^{5}$ nuclear spins, we use the time evolution of the distribution function to calculate the non-equilibrium dynamics of the 
correlation function $\esziz$ for two different electron spin temperatures and
different ratios $\we0/\wn0$.
The time-dependent expectation value is obtained from Eq.~\eqref{eq:expect} assuming that the off-diagonal contributions in the density operator can be neglected.
As initial condition for the electron-nuclear spin system, we start with the completely disordered state, i.e., with an occupation of the spin states in the distribution function according to their degeneracy 
\begin{align}
g_0(\iz,\sz) = 2^{-N-1} {N \choose N_+(\iz) } ,
\end{align}
where $N_+(\iz) = N/2+\iz$ and ${a\choose b}$ is the binomial coefficient.

\begin{figure}[t!]
\centering
\includegraphics[scale=1]{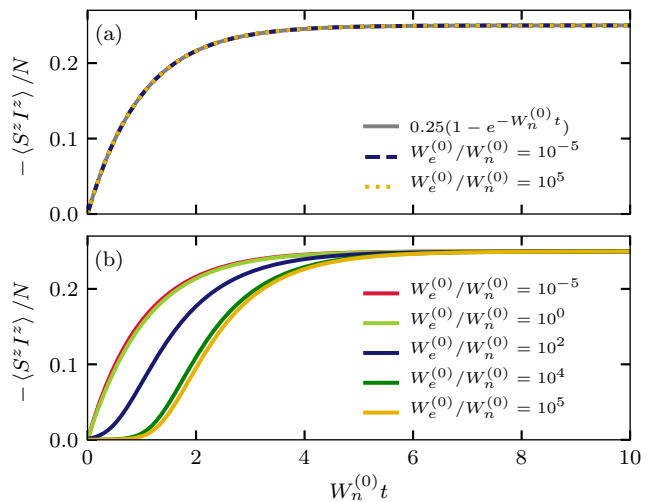}
\caption{Formation of the nuclear polaron ($N=10^5$, $\bn \omega_h =10^4$) at different inverse electron temperatures, (a) $\be \omega_h = 5$ and (b) $\be \omega_h = 0.05$. The time axis is scaled by the nuclear spin-flip rate $\wn0$.}
\label{fig:timeevol}
\end{figure}

For temperatures well below the transition temperature, $\bn \omega_h =10^4$ and $\be \omega_h = 5$, the correlation between electron and nuclei evolves on a time scale determined by the nuclear spin-flip rate, 
see Fig.~\ref{fig:timeevol}(a).
Introducing the dimensionless time $\tau=\wn0 t$ reveals that the evolution of $\esziz$ does not depend on the electron flip rate at low temperatures.
The evolution for various ratios $\we0 / \wn0$ follows a universal function independent of whether the electron or the nuclear spin flips occur with a faster rate.
This universal function is given by
$\esziz/N = -(1-\exp (-\tau)) / 4$ (indicated by the gray line)
where only $\wn0$ enters via $\tau$.

This independence on the electron spin-flip rate $W_e^{(0)}$ indicates that the Overhauser field polarization suppresses electron spin-flip processes.
While a single nuclear spin flip only involves a small change in the total energy of the coupled system, the electron spin flip induces a large energy change that is exponentially suppressed as stated in Eq.~\eqref{eq:we}.
The state of maximum anti-correlation can be reached without any electron spin flip while at least half of the nuclear spins have to flip starting from the disordered state.
As a consequence the time scale, on which $\esziz$ builds up, is solely given by the nuclear flip rate $W_n^{(0)}$.

Increasing the electron spin temperature, $\be \omega_h = 0.05$, see Fig.~\ref{fig:timeevol}(b), the time evolution of $\esziz$ changes.
While at the end still the maximum anti-correlation is reached, 
the buildup differs from a purely exponential growth since the electron flip rate 
gains influence according to Eq.~\eqref{eq:we}.
For a fast electron flip rate, the electron spin flips disturb the formation of the polaron state.
Reorientation of the electron when the nuclear spins have started to align according to the momentarily electron state hinders the polaron formation.
Thus, the polaron forms more slowly in comparison to lower temperatures.
After some time these random electron spin flips are suppressed as the nuclear spins have managed to build an Overhauser field strong enough to prevent electron spin flips.
Then, again, the further evolution of the polaron state depends on the nuclear spin-flip rate.
The effect of electron spin flips is especially strong for large $\we0$, however, when the electron already follows the nuclear spins almost instantaneously a further increase of $\we0$ does not alter the dynamics much anymore.
In the opposite case, where the electron spin-flip rate is comparable to the nuclear rate or slower, we recover the exponential behavior depending on $\wn0$.

\section{Conclusion}
\label{sec:conclusion}

We derived a set of kinetic equations to describe the spin dynamics and, particularly, the formation of a nuclear-polaron state in an electron-nuclear spin system.
Such a polaron state may occur under optical cooling of the nuclear spin system in semiconductors with localized charge carriers, such as bulk materials with donor-bound electrons or quantum dot structures.
Our theory incorporates the electron coupling to the bath of nuclear spins within the central spin model, as well as electron and nuclear spin flips due to the exchange with two distinct reservoirs of temperatures $T_e$ and $T_n$.
Phenomenological temperature-dependent flip rates for electron and nuclear spins are introduced. 
For the analytical and numerical study of the spin system's expectation values we focused on a model Hamiltonian for the hyperfine interaction which is of Ising-type, however the generalization to the isotropic central spin model would provide an advance towards real semiconductor systems and remains a topic for future investigations.

The polaron formation as a function of the electron and nuclear spin temperatures is studied by analyzing the electron-nuclear spin correlation function and the nuclear distribution function.
We discuss the criteria of the polaron formation and it turns out that a relatively sharp transition line can be determined by the condition where the nuclear spin distribution function demonstrates two peaks at $I^z \ne 0$ instead of one at $I^z=0$.
This temperature line translates to a pronounced peak in the fluctuations of the absolute value of the total nuclear spin and agrees with the mean-field critical temperature in a wide range of parameters.
Generally, due to the finite size of the studied system, the nuclear-polaron formation is described by a crossover rather than by a phase transition.

We also investigate the role of the distribution of the hyperfine-coupling constants and show that it can be semi-quantitatively taken into account by reducing the effective number of nuclei interacting with the electron spin in the box model.

Furthermore, the derived kinetic equations yield the temporal evolution of the polaron state.
We find that the polaron formation is governed by spin flips in the nuclear spin bath.
As a result, the dynamics in the cooled system depends mostly on the nuclear flip rate
while the electron flip rate has less effect on the time scale of the polaron formation.
The developed formalism allows one to address the fluctuations in the course of polaron formation.
The study of the temporal fluctuations of the electron and nuclear spins and their cross-correlations in the regime of nuclear-polaron formation is a task for future.


\begin{acknowledgments}
We are grateful to K. V. Kavokin and D. S. Smirnov for discussions.
We acknowledge financial support by the Deutsche Forschungsgemeinschaft and the Russian Foundation of Basic Research through the transregio TRR 160 within the Projects No.\ A4, and No.\ A7.
M.M.G. was partially supported by RFBR-DFG project No. 19-52-12038.
The authors gratefully acknowledge the computing time granted by the John von Neumann Institute for Computing (NIC) under Project HDO09 and provided on the supercomputer JUWELS at the J\"ulich Supercomputing Centre.
\end{acknowledgments}


%

\end{document}